\documentclass[12pt]{article}
\usepackage[english]{babel}
\usepackage{epsfig}
\usepackage{graphicx}
\usepackage{longtable}
\usepackage{amstext}
\usepackage{amssymb}
\sloppypar
\usepackage[font=small,labelfont=bf,labelsep=period]{caption}
\begin{document}
\begin{center}
{\Large\bf Comprehensive Study of the Magnetic Stars HD 5797 and HD 40711 with Large Chromium and Iron Overabundances} \\
\vspace{1cm}
{\bf E. A. Semenko\footnote{sea@sao.ru}, I. A. Yakunin, and E. Yu. Kuchaeva}
\end{center}
{\it Special Astrophysical Observatory, Russian Academy of Sciences, Nizhnii Arkhyz, Karachai-Cherkessian Republic, 369167 Russia}

\vspace{1cm}

{\bf Abstract~---} We present the results of a comprehensive study of the chemically peculiar stars HD 5797 and HD 40711. The stars have the same effective temperature, $T_\textrm{eff} = 8900$\,K, and a similar chemical composition with large iron~($+1.5$ dex) and chromium~($+3$ dex) overabundances compared to the Sun. The overabundance of rare-earth elements typically reaches $+3$ dex. We have measured the magnetic field of HD 5797. The longitudinal field component $B_\mathrm{e}$ has been found to vary sinusoidally between $-100$ and $+1000$\,G with a period of 69 days. Our estimate of the evolutionary status of the stars suggests that HD 5797 and HD 40711, old objects with an age $t\approx5\cdot10^8$\,yr, are near the end of the core hydrogen burning phase.


\section{INTRODUCTION}
About 15\% of the hot stars in the upper main sequence of the Hertzsprung?Russell diagram have pronounced chemical anomalies. The spectra of these stars exhibit unusually strong or weak lines of iron, chromium, silicon, and several other chemical elements, including rare earths. Following Preston~(1974), such objects are called chemically peculiar~(CP) stars. Spectropolarimetric methods of observations of chemically peculiar stars reveal strong magnetic fields in their atmospheres whose strength varies over a wide range from tens of gauss~(Auriere et al. 2007) to tens of kilogauss. Data on the fields of about 350 objects have been obtained in more than 60 years of studies of the effect of magnetic stars~(Romanyuk and Kudryavtsev 2008); these are mostly the measurements of the longitudinal magnetic field component $B_\mathrm{e}$ averaged over the visible stellar hemisphere. The lower limit of $B_\mathrm{e}$ for CP stars is determined by the accuracy of the applied Zeeman measurement methods.

The studies of various authors show that the distribution of magnetic stars, as a subclass of CP stars, on the main sequence is relatively uniform~(see, e.g., Kochukhov and Bagnulo 2006; Landstreet et al.2007): there are both stars leaving the main sequence and those preparing to reach it. Thus, by observing CP stars of various ages, we can study the evolution of their chemical composition and magnetic fields. In addition, the question of the interrelation between the magnetic field strength, geometry, and the pattern of chemical anomalies remains topical.

In their paper aimed at studying the chemical stratification in the atmosphere of the star HD 133792, Kochukhov et al.~(2006) suggested separating a special group of evolved CP stars with very large chromium and iron overabundances. Anomalously weak, for peculiar stars of the same temperature range, lines of rare-earth elements are a characteristic feature of such objects. Weak magnetic fields are another common feature of the new group of stars. HD 133792 and HD 204411 are previously studied representatives of evolved magnetic stars~(Ryabchikova et al. 2005). Quite recently, this list has been expanded: while studying HD 103498 for rapid pulsations, Joshi et al.~(2010) performed a detailed analysis of its chemical composition. In addition, the authors managed to estimate the positions of HD 133792, HD 204411, and HD 103498 on the Hertzsprung?Russell diagram, whence it followed that all of the stars under consideration either have already passed the core hydrogen burning phase or are close to its end.

In this paper, we present the results of our spectroscopic analysis of two magnetic CP stars with unusually strong iron and chromium lines. The spectra of HD 5797 and HD 40711 look very similar~(Fig. 1), and the two stars are approximately at the same distance from us. Besides, our comparison of the spectra of the two stars to parts of the spectrum of HD 133792 provided by T.A. Ryabchikova revealed several features in common.

The stars HD 5797 and HD 40711 have become objects of observations on the 6-m Special Astro-physical Observatory~(SAO) telescope as part of our program of research on CP stars whose magnetic fields were first detected at SAO. Preference was given to stars with sharp lines and with known distances.

The first direct magnetic field measurements for HD 5797 were presented by Iliev et al.~(1992). They performed photographic observations using the Main stellar spectrograph of the 6-m telescope and one Zeeman spectrum was taken at that time. Using two different methods for estimating the longitudinal magnetic field, the authors found $B_\mathrm{e}$ to be $-2200$\,G with an error of 300\,G. Still earlier, the magnetic properties of the star were studied by Preston~(1971) who estimated the surface field to be 1.8\,kG from the differential broadening of spectral lines. To refine information on the magnetic field strength and variability, we performed new observations on the 6-m telescope using a circular polarization analyzer. The results of this work will be described below.

Adelman~(1973) studied the chemical composition of HD 5797, pointing out that it differs from other objects of the same spectral type by large Fe, Cr, Ti, Mn, Nd, Gd, and Si overabundances, while the Sr, Y, Ca, and V abundances are solar. In addition, the photometric and spectroscopic variability of the star was studied previously~(Wolff 1975; Barzova and
Iliev 1988). According to the results of studying the variability of spectral line equivalent widths presented by Barzova and Iliev~(1988), there are no strong Eu~\textsc{ii} lines in the spectrum reported by Preston~(1971) and Adelman~(1973).The equivalent widths of Fe~\textsc{i}--\textsc{ii}, Cr~\textsc{i}--\textsc{ii}, Ti~\textsc{ii}, and Sr~\textsc{ii} lines change by a factor of 1.5--2 with the 69-day period suggested by Wolff~(1975).

In contrast to HD 5797, the star HD 40711 has been studied less actively. Its \textit{uvby}~(Vogt and Faundez 1979; Masana et al. 1998) and $\Delta a$~(Maitzen and Vogt 1983; Paunzen et al. 2005) photometry is available. It is the large $\Delta a$ index that was the reason why HD 40711 was included in the program of searches for new magnetic stars. As a result, El'kin et al.~(2003) measured four Zeeman spectra taken on the 6-m SAO telescope and found the star's longitudinal magnetic field to vary within the range from $-650$ to $+330$\,G, but they failed to find the variability period.

Carrier et al.~(2002) reported that HD 40711 is an SB1 spectroscopic binary with an orbital period of $1245^\mathrm{d}$ and a high orbital eccentricity, $e=0.8$. The same paper provides estimates of the star's rotational velocity~($v_\mathrm{e}\sin i=2$\,km s$^{-1}$) and effective temperature~($T_\mathrm{eff}=9328$ K).

Since no detailed spectroscopic analysis of the star HD 40711 has been performed previously, we have decided to fill in this gap and to include it in our program of observations with the echelle spectrometer of the 6-m telescope.

\section{OBSERVATIONAL DATA AND THEIR REDUCTION}
High-resolution spectroscopy was obtained with the NES echelle spectrometer mounted at the Nasmyth-2 focus of the 6-m SAO telescope~(Panchuk et al. 2009). We took a total of four echelle spectra for HD 5797 in the wavelength range 4600--6780\,\AA\AA.  To study the star HD 40711, we used two echelle spectra taken by D.O. Kudryavtsev in October 2007 using the same equipment with a similar configuration. In all cases, the spectral resolution $\lambda/\Delta\lambda$ was close to 40 000 and the signal-to-noise ratio in the $\lambda=5500$\,\AA\ region was at least 150.

The \textsc{Reduce} software package~(Piskunov and Valenti 2002) was used for the primary reduction of 2D images. The reduction was performed according to a standard scheme and included the following steps: CCD bias subtraction, flat fielding, stray light subtraction, and extraction of 1D spectra. Using the \textsc{Wavecal} procedure from the same software package, we passed from the CCD coordinate system to wavelengths. As a reference, we used the spectrum of a ThAr lamp. Subsequently, the individual echelle orders were normalized to the continuum level using the task \textsc{Continuum} from the IRAF data reduction package. To reduce the errors of the continuum placement near hydrogen lines, we attempted to fit the blaze function in a direction across the dispersion and to perform the normalization based on this estimate. Applying this procedure allows the location of the continuum near $H\alpha$ to be reliably determined, while the accuracy of the continuum placement near $H\beta$ is considerably lower.

We took five Zeeman spectra of HD 5797 with the Main stellar spectrograph~(MSS) of the 6-m telescope. Two more spectra were taken at our request on the same equipment by D.O. Kudryavtsev and G.A. Chountonov. This spectrograph is equipped with a circular polarization analyzer with an image slicer and a rotating quarter-wave plate~(Chountonov 2004). We reduced the spectra and measured the longitudinal field using programs from the \textsc{Zeeman} package written by D.O. Kudryavtsev for the ESO MIDAS system. The spectrum of a hollow-cathode ThAr lamp was used as a comparison one. Similar to the echelle spectra, we also normalized the polarization spectra using the \textsc{Continuum} task from the IRAF package.

Table 1 presents all information about our spectra, including the HJD time, the spectral resolution, the spectral range, and the signal-to-noise ratio. In addition, for the star HD 5797, we give the phase derived from the ephemeris presented by Wolff~(1975) and the measured longitudinal magnetic field strength.

\section{PHYSICAL PROPERTIES OF THE STARS HD 5797 AND HD 40711}
\subsection*{\it Stellar Effective Temperature, Surface Gravity, and Rotation}

It is not easy to determine the physical characteristics of CP stars. An improper or incomplete allowance for such factors as the magnetic line broadening, the energy redistribution in the continuum, the interstellar extinction, etc can be of crucial importance.

According to the results of a new HIPPARCOS data reduction~(van Leeuwen 2007), the stars under consideration are fairly distant objects: the parallaxes of HD 5797 and HD 40711 are $2.19\pm0.87$ and $2.66\pm0.95$ mas, respectively. The spectra of both stars exhibit strong sodium doublet lines of an interstellar origin, which is also indicative of a great distance. However, the sodium lines are separated in HD 40711 due to its radial velocity, while both D1 Na lines in the spectrum of HD 5797 merge together and it is virtually impossible to separate them. We used two methods to determine the color excess $E(B-V)$. According to the first estimate based on the interstellar reddening maps from Lucke~(1978), $E(B-V)$ is $0^{m}\!\!.17$ for HD 5797 and $0^{m}\!\!.09$ for HD 40711. The second method is based on the relation between $E(B-V)$ and the equivalent width $W_\mathrm{eq}(\textrm{D1})$ of the interstellar sodium D1 doublet from Munari and Zwitter~(1997). When measuring the equivalent width for HD 40711, we used the right component of the line belonging to the interstellar medium. As a result, the second method yields results in excellent agreement with those obtained from the maps by Lucke~(1978).

We determined the effective temperature and surface gravity in two stages. At the first stage, we used the available calibrations of the indices of the Str\"omgren and Geneva photometric systems taken from the Geneva GCPD database~(Mermilliod et al.1997). Applying the calibrations from Napiwotzki et al.~(1993) separately for the Str\"omgren $[u-b]$ and $[c1]$ indices gave a discrepancy in effective temperatures by almost 500 K. We obtained $T[u-b]=9060$\,K, $T[c1]=9550$\,K, and $\log g=3.69$ for HD 5797. The corresponding parameters for HD 40711 turned out to be 9084\,K, 9519\,K, and 3.75.

We used the calibrations from K\"unzli et al.~(1997) to determine $T_\mathrm{eff}$ and $\log g$ from the Geneva photometric indices. The parameters $pT$ and $pG$, which are well suited for determining the effective temperature and surface gravity in stellar atmospheres in the range $T_\mathrm{eff}=8500-10500$\,K, are sensitive to interstellar reddening. Taking $E(B2-V1)$ to be $0^{m}\!\!.15$ for HD 5797, we found $T(pT, pG)=8910$\,K, $\log g=3.65$ for metallicity $[M/H]=0$ and $T(pT,pG)=8743$\,K, $\log g=3.23$ for $[M/H]=+1$. Similarly, the combinations of parameters found for HD 40711 ($E(B2-V1)=0^{m}\!\!.08$) are $T(pT,pG)=9040$\,K, $\log g=3.85$, $[M/H]=0$ and $T(pT,pG)=8831$\,K, $\log g=3.44$, $[M/H]=+1$.

The observed agreement between the estimates obtained from the same indices can be explained by the remarkable similarity between both stars: not only their photometric indices and distances almost coincide, but also the appearance of their spectra indicates that the stars should have virtually identical characteristics.

At the second stage, we compared the synthetic hydrogen line profiles computed for various combinations of $T_\mathrm{eff}$, $\log g$, and $[M/H]$ and found that the best agreement was achieved for an effective temperature of 8900\,K and a logarithm of the surface gravity of 3.4 at a metallicity $[M/H]$ close to $+1$. Figure 2 presents the H$\alpha$ and H$\beta$ hydrogen lines in the spectrum of HD 5797 with the synthetic profiles superimposed on them. A similar comparison was also made for the H$\beta$ line in the spectrum of HD 40711. The atmospheric parameters of the program stars derived from the hydrogen lines agree well with those determined from the calibrations of the Geneva photometric system. A lower temperature than that inferred from the calibrations of the Str\"omgren photometry is suggested, for example, by the presence of a considerable number of neutral titanium lines in the spectra of the program stars. Thus, below we adopt $T_\mathrm{eff}=8900\pm150$\,K and $\log g=3.4\pm0.1$\,dex.

The sharp line profiles in our high-resolution spectra are indicative of a slow rotation of the program stars. To estimate $v\sin i$, in each case, we used the two Fe\,\textsc{i} $\lambda5434.52$\AA\ and $\lambda5576.09$\AA\ lines insensitive to the magnetic field. The FWHMs of these lines turned out to be determined by the instrumental profile of the NES spectrograph, i.e. the stars HD 5797 and HD 40711 rotate with a velocity
$v\sin i \leq 7$\,km$\:$s$^{-1}$. This estimate is consistent with the data on the rotation of the program stars that can be found in Carrier et al.~(2002). These authors give $v\sin i = 3$\,km$\:$s$^{-1}$ for HD 5797 and $v\sin i = 2$\,km$\:$s$^{-1}$ for HD 40711.

\subsection*{\it Evolutionary Status}
We determined the evolutionary status of the stars from their positions on the theoretical temperature-luminosity diagram. Using the formula
$$\log \frac{L}{L_\odot}=-\frac{M_\mathrm{v}+BC-M_\mathrm{bol}(\odot)}{2.5}$$
we found the luminosities of the program stars. The absolute magnitude is determined here as follows:
$$M_\mathrm{v}=m_\mathrm{v}+5+5\,\log \pi - A_\mathrm{v}.$$

According to Flower~(1996), the bolometric correction $BC$ for a temperature of 8900\,K is $-0.009$. The transition from the color excess to the visual extinction is made via the relation $A_\mathrm{v} = 3.1\:E(B-V)$. Thus, the luminosity $\log(L/L_{\odot})$ of HD 5797 is $2.05\pm0.20$, while the luminosity of HD 40711 is slightly lower: $\log(L/L_{\odot})=1.75\pm0.20$. The low accuracy of the parallax determination makes a major contribution to the error in the luminosity.

Comparison of the luminosity and temperature data with the evolutionary tracks from Girardi et al.~(2000) in Fig.~3 suggests that both objects have gone far away from the zero-age main sequence and that HD 5797 is close to the end of the core hydrogen burning phase. Our stars are very similar in this characteristic to other representatives of evolved magnetic stars. In addition, because of the large error in the luminosity, it is difficult to accurately estimate the stellar masses; we can only say that both objects are fairly massive ($M/M_{\odot}\approx2.5-3.0)$, while comparison with isochrones allows the ages of the stars to be estimated: $t=450-630$\,Myr.

\subsection*{\it The Magnetic Field of HD 5797}
Until recently, it was difficult to reach an unambiguous conclusion about the magnetic properties of HD 5797 due to the lack of observational data. The strong magnetic field, greater than $-2$ kG, found by Iliev et al.~(1992), strongly suggests that the star belongs to the class of magnetic stars. However, to ascertain the field strength and variability pattern, we decided to perform additional spectropolarimetric observations on the 6-m telescope. Table 1 presents the results of our six measurements of the longitudinal magnetic field; one more measured Be was kindly provided to us by D.O.~Kudryavtsev. Since all of the Zeeman spectra listed in the table were taken and reduced using the same equipment and the same software packages, we can assert that the presented data are homogeneous. In Fig.~4, the magnetic field component $B_\mathrm{e}$ is plotted against the rotation phase. According to our measurements, HD 5797 has a variable magnetic field; $B_\mathrm{e}$ varies within the range from $-100$ to 1000~G and cannot reach $-2$\,kG, as was pointed previously by Iliev et al.~(1992), especially since the abundance of narrow lines allows accurate magnetic field measurements to be made, and the measurement error of $B_\mathrm{e}$ is about 50\,G in our case. We do not rule out the possibility that additional observations will be required to refine the magnetic field variability period, but it can be concluded already now that the 69-day photometric period suggested by Wolff~(1975) describes our results satisfactorily.

Despite the fairly high spectral resolution and low rotational velocities, the surface magnetic field of HD 5797 is so weak that it does not cause any visible Zeeman splitting or at least broadening of magnetically sensitive lines. Therefore, based on the available $B_\mathrm{e}$ measurements and assuming a dipole field structure, we concluded that the magnitude of the surface magnetic field strength for the star does not exceed 3 kG.

\subsection*{\it Chemical Composition of the Stars}
We analyzed the chemical composition of both stars by the method of theoretical curves of growth in the form implemented in Kurucz's \textsc{Width9} code modified by V.~Tsymbal for data input in the VALD format~(Kupka et al. 1999). Our choice of the method was dictated by the presence of many narrow, weakly blended lines in the spectrum and a weak magnetic field. For our measurements, we used only the high-resolution spectra taken with the NES echelle spectrograph. To identify the lines of individual elements, we computed a synthetic spectrum using the \textsc{Synth3} code written by N.E.~Piskunov and subsequently improved by Kochukhov~(2007).  This version of the code works in the LTE approximation and is optimized for the use of input atomic data in the VALD format. The model atmospheres were computed using the \textsc{Atlas9} code~(Kurucz 1993) with a scaled solar metallicity. A good overlap of the spectra for HD 5797 in wavelengths allowed us to analyze the abundances of many elements. Although all spectra were taken on different dates, comparison of the equivalent widths of individual lines measured on neighboring dates~(Fig.~5) suggests the absence of significant spectroscopic variability on this time scale, which also indirectly points to a long variability period of the star. In addition to the three high-resolution spectra taken in August 2008~($\phi\approx0.57$), we obtained one spectrum in September 2009 near the maximum of the curve of the longitudinal magnetic field.

We took into account the magnetic field effects in our computations of the synthetic spectra by introducing a pseudo-microturbulence determined from the absence of a relation between the equivalent widths of iron lines and the abundance derived from them. The value of $\xi_\mathrm{micro}$ is 0.8 km$\:$s$^{-1}$ for the spectra of HD 40711, while for HD 5797 it is 0.7 km$\:$s$^{-1}$ at phases corresponding to a weak field $B_\mathrm{e}$ and 1.55 km$\:$s$^{-1}$ when the longitudinal field is at a maximum.

The results of our abundance determinations are summarized in Table 2. For comparison, apart from the solar composition, the table gives the abundances of chemical elements in the atmosphere of the star HD 103498 with a temperature $T_\mathrm{eff}=9500$\,K~(Joshi et al. 2010), which is a representative of the already studied group of evolved CP stars.

Having analyzed Table 2, we can note that HD 5797 and HD 40711 are characterized by approximately the same degree of anomalies. Despite the small number of identified silicon lines, we may conclude that this element is slightly overabundant compared to the Sun. The Si abundance slightly increases with decreasing rotation phase of HD 5797, but it is unreasonable to talk about the reliability of this conclusion. We managed to find a small number of calcium lines in the spectra of the program objects; the calcium abundance turned out to be nearly solar. Titanium is considerably overabundant in both stars and, in addition, its abundance slightly increases with $B_\mathrm{e}$. Note the large spread in the titanium abundance determined from individual Ti\,\textsc{ii} lines. As we see from Table 2, the manganese abundance in the atmosphere of HD 5797 increases with longitudinal field. This conclusion is confirmed by a considerable enhancement of the Mn lines in the spectrum at phase $\phi=0.17$. Manganese is represented less significantly in the spectrum of HD 40711, but its overabundance compared to the Sun reaches 1.3\,dex even in this case. In both stars, iron and chromium exhibit an overabundance of 1.5 and 3\,dex, respectively. Both program stars are very similar in this characteristic
to HD 103498. However, in contrast to the latter, HD 5797 has a large strontium overabundance. The abundance of this element is higher than that on the Sun by more than 2.5\,dex. In the working range of the available data, we failed to identify any strontium lines in the spectrum of HD 40711. Thus, the assumption by Joshi et al.~(2010) about the influence of stellar rotation on the strontium abundance in the observed atmospheric layers may have physical grounds, because the stars we studied belong to very slow rotators. Almost all rare-earth elements exhibit a slight overabundance compared to the Sun, their lines are weak and few in number; therefore, the results of rare-earth abundance determinations are very sensitive to the continuum placement errors.

\section{DISCUSSION AND CONCLUSIONS}
In this paper, we carried out a comprehensive study of two chemically peculiar stars with weak magnetic fields. Our goal was to estimate the chemical composition and to measure the dependence of the magnetic field on rotation phase.

Our long-term spectrophotometric observations on the 6-m SAO telescope have revealed that the longitudinal magnetic field of the star HD 5797 varies sinusoidally between $-100$\,G and 1\,kG with the 69-day period suggested by Wolff~(1975).

A chemical abundance analysis confirmed our initial assumption that both stars exhibit the greatest abundance anomalies for iron~($+1.6$ dex compared to the Sun), chromium~($+3$\,dex), and strontium~($+3$ dex). We managed to identify the lines of the latter element only in the spectrum of HD 5797. The rare-earth abundances were found to be slightly below the mean for unevolved stars of the same temperature~--- the overabundance compared to the Sun is 2~-- 3 dex. No Nd\,\textsc{ii}$/$Nd\,\textsc{iii} ionization disequilibrium, a well-known rare-earth anomaly characteristic of most cooler oscillating~(roAp) chemically peculiar stars~(Ryabchikova et al. 2004), is observed.

In future, it seems interesting to perform a stratification analysis of the distribution of chemical elements in the atmospheres of HD 5797 and HD 40711 and to compare the results with those obtained previously for the stars HD 133792 and HD 204411, which are characterized by a similar chemical composition, have similar temperatures, and, most importantly, belong to old representatives of magnetic stars, just as the stars we studied. It is possibly the evolutionary status of the listed stars and the strength of their magnetic fields that are responsible for the observed chemical anomalies. If the magnetic flux is assumed to change insignificantly in the main-sequence lifetime of the star, then the observed weak magnetic fields of evolved stars should result from their expansion as hydrogen will be depleted in their interiors. Our simplified computations of the stellar evolution using the \textsc{Evolving ZAMS} code~(Paxton 2004) show that by the time a star with an initial mass of 2.9 $M_{\odot}$ leaves the main sequence, its sizes increase by more than a factor of 2 compared to those on the zero-age main sequence. Therefore, we believe that in the rarefied atmospheres of evolved stars, one may expect an increase in the importance of the magnetic field in the matter diffusion processes. Further accumulation of observational data and their analysis are needed to clarify the situation. A theoretical study of the question is of no lesser importance. So far, one thing is clear~--- stars with weak magnetic fields whose evolutionary status is similar to that of HD 5797 and HD 40711 exhibit large overabundances of Cr, Fe, and some other elements, while the rare-earth elements are represented poorly. At the same time, we know stars in the temperature range 8500~-- 9500\,K with strong magnetic fields and prominent rare-earth lines in their spectra. As examples of such stars, we can mention HD 66318~(Bagnulo et al.~2003) and HD 158450~(Jilinski et al.~2009), but they are both young objects.

Since the stars we studied represent, in a sense, the extreme case of the overabundances of some chemical elements, note that when the effective temperature is determined by fitting the hydrogen lines, the H$\beta$ line profile in the wings and when passing from the wings to the core can be described satisfactorily only for an enhanced metallicity. The hydrogen line cores themselves are described poorly. In future, the use of model atmospheres for HD 40711 and HD 5797 with allowance made for their individual abundances~(see, e.g., Shulyak et al. 2008) may improve the agreement between the theoretical and observed hydrogen line profiles, but satisfactory results can be achieved even when using models with a scaled solar  metallicity.

The question about the magnetic field contribution to the formation of depressions in the continuum spectra of HD 5797 and HD 40711 also remains open. Both stars are known to have a significant depression in the continuum at 5200\,\AA: the $\Delta a$ index in the Vienna photometric system is 0.038 for HD 5797 and 0.043 for HD 40711. Taking into account the weak general magnetic field, we believe that the observed anomaly in the energy distribution is primarily attributable to the iron and chromium abundances, because the remaining elements with large overabundances are represented considerably more poorly in the spectrum. Many iron lines that can be responsible for the observed anomalies in the continuum energy distribution are observed in the spectra of our program stars in the region of the depression. The magnetic field plays a secondary role in such stars.

In conclusion, note that, although the available set of model atmospheres and codes for computing synthetic spectra yields satisfactory results when describing old magnetic stars with extremely high abundances of individual chemical elements, it is necessary to develop self-consistent model atmospheres that would take into account the influence of the magnetic field. This can be an important step toward understanding the evolution of CP stars with age.

\section*{ACKNOWLEDGMENTS}
We are grateful to D.O.~Kudryavtsev and G.A.~Chountonov for their help in obtaining the observational data. We also thank I.I.~Romanyuk and T.A.~Ryabchikova for a helpful discussion of the results and a number of remarks that improved the content of the paper. This work was supported in part by the Russian Foundation for Basic Research~(projects nos. 06-02-16110-a and 09-02-00002-a).

\textit{Translated by N. Samus'}

\clearpage
\begin{table}[t]
\vspace{6mm}
\centering
{{\bf Table 1.} Information about the spectra used here.}\label{log}
{\scriptsize
\vspace{5mm}\begin{tabular}{r c  c c c  c c c}\hline\hline 
  Star & HJD, 2450000+ & Instrument & $S/N$ & $\Delta\lambda$, \AA & $\lambda/\Delta\lambda$ & $B_\mathrm{e}$, $\sigma$, G & Phase \\
\hline
HD 5797   & 4698.5147 & NES   & 150  & 4600--6090 & 40000 &            & 0.54 \\
       & 4700.5148 & NES   & 160  & 5330--6780 & 40000 &            & 0.57 \\
       & 4702.5150 & NES   & 160  & 4700--6160 & 39000 &            & 0.60 \\
       & 4750.3506 & MSS  & 300  & 4414--4654 & 14500 & $810\pm50$ & 0.29 \\
       & 4783.4183 & MSS  & 300  & 4215--4455 & 14500 & $0\pm45$   & 0.77 \\
       & 4955.5003 & MSS  & 300  & 4402--4642 & 14500 & $920\pm50$ & 0.27 \\
       & 5077.5176 & MSS  & 300  & 4762--5000 & 14500 & $830\pm50$ & 0.04 \\
       & 5086.5438 & NES   & 160  & 4800--6250 & 41000 &            & 0.17 \\
       & 5171.4115 & MSS  & 300  & 4398--4636 & 14500 & $660\pm50$ & 0.40 \\
       & 5197.1493 & MSS  & 200  & 4316--4556 & 14500 & $-100\pm50$& 0.76 \\
       & 5255.2172 & MSS  & 300  & 4360--4600 & 14500 & $-65\pm45$ & 0.61 \\
\hline
HD 40711  & 4395.5857 & NES   & 200  & 4690--6168 & 42500 & & \\
       & 4396.5248 & NES   & 200  & 4690--6168 & 40000 & & \\
\hline
\end{tabular}}
\end{table}

\clearpage
\begin{table}[t]
{{\bf Table 2.} Elemental abundances in the atmospheres of the program stars. For comparison, the elemental abundances in the atmospheres of HD 103498 and the Sun are given. $n$ is the number of measured lines.}\label{abund}
\centering
\vspace{5mm}{\footnotesize\begin{tabular}{l | c c | c c | c c | c | c}
\hline\hline
     & \multicolumn{4}{| c |}{HD 5797} & \multicolumn{2}{c |}{HD 40711} & HD 103498 & Sun \\
\cline{2-8}
     & \multicolumn{2}{| c |}{$\phi=0.57$} & \multicolumn{2}{c |}{$\phi=0.17$} & & & &  \\
\cline{2-5}
Species  & {$\log N/N_\mathrm{tot}$~($\sigma$)} & $n$ & {$\log N/N_\mathrm{tot}$~($\sigma$)} & $n$ & {$\log N/N_\mathrm{tot}$~($\sigma$)} & $n$ & {$\log N/N_\mathrm{tot}$} & {$\log N/N_\mathrm{tot}$} \\
\hline
Na I  & $-5.66$ (0.16) & 2   & $-5.49$ (:)    & 1   & $-5.19$ (:)    & 1   & $-5.09$  & $-5.87$ \\
Mg I  & $-3.49$ (0.19) & 5   & $-3.14$ (0.14) & 3   & $-3.48$ (0.22) & 3   & $-3.58$  & $-4.51$ \\
Mg II & $-3.55$ (:)    & 1   & $-3.80$ (:)    & 1   & $-3.53$ (:)    & 1   & $-4.43$  & $-4.51$ \\
Si I  & $-4.07$ (0.01) & 2   & $-3.78$ (0.06) & 2   & $-3.97$ (0.02) & 3   & $-3.65$  & $-4.53$ \\
Si II & $-3.97$ (:)    & 1   & $-3.85$ (:)    & 1   & $-3.78$ (:)    & 1   & $-3.64$  & $-4.53$ \\
Ca I  & $-5.54$ (:)    & 1   &                &     &                &     &          & $-5.73$ \\
Ca II & $-6.08$ (:)    & 1   &                &     & $-5.63$ (:)    & 1   & $-5.91$  & $-5.73$ \\
Sc II & $-8.70$ (0.04) & 3   & $-8.72$ (:)    & 1   & $-9.06$ (0.19) & 3   &          & $-8.99$ \\
Ti I  & $-5.26$ (0.15) & 10  & $-4.78$ (0.19) & 9   & $-5.06$ (0.07) & 5   &          & $-7.14$ \\
Ti II & $-5.35$ (0.20) & 10  & $-4.82$ (0.31) & 18  & $-5.05$ (0.23) & 30  & $-6.45$  & $-7.14$ \\
Cr I  & $-3.41$ (0.16) & 63  & $-3.25$ (0.24) & 35  & $-3.31$ (0.25) & 70  & $-3.25$  & $-6.40$ \\
Cr II & $-3.29$ (0.19) & 81  & $-3.03$ (0.25) & 85  & $-3.15$ (0.26) & 149 & $-3.31$  & $-6.40$ \\
Mn I  & $-4.91$ (0.14) & 7   & $-4.27$ (0.26) & 3   & $-5.40$ (0.22) & 6   & $-5.94$  & $-6.65$ \\
Mn II & $-4.70$ (0.23) & 8   & $-4.24$ (0.15) & 12  & $-5.33$ (0.13) & 5   & $-5.72$  & $-6.65$ \\
Fe I  & $-2.99$ (0.25) & 114 & $-3.03$ (0.26) & 80  & $-2.99$ (0.22) & 124 & $-2.98$  & $-4.59$ \\
Fe II & $-2.78$ (0.23) & 117 & $-2.73$ (0.25) & 81  & $-2.87$ (0.25) & 143 & $-3.01$  & $-4.59$ \\
Ni I  & $-5.79$ (0.14) & 4   & $-5.54$ (0.18) & 3   & $-5.66$ (0.22) & 4   & $-5.34$  & $-5.81$ \\
Sr I  & $-6.17$ (0.11) & 4   &                &     &                &     &          & $-9.12$ \\
Sr II & $-6.56$ (0.21) & 3   & $-6.32$ (:)    & 1   &                &     & $\leq-8.5$ & $-9.12$ \\
Y II  & $-9.42$ (0.36) & 2   & $-9.26$ (0.21) & 3   & $-9.27$ (0.35) & 2   &          & $-9.83$ \\
Zr II & $-8.53$ (0.07) & 2   & $-8.53$ (0.03) & 2   &                &     &          & $-9.45$ \\
Ba II & $-9.32$ (:)    & 1   & $-9.51$ (:)    & 1   &                &     & $-8.64$  & $-9.87$ \\
La II & $-8.68$ (0.42) & 5   & $-8.52$ (0.33) & 2   & $-8.75$ (0.02) & 2   &          & $-10.91$ \\
Ce II & $-8.42$ (:)    & 1   & $-8.20$ (:)    & 1   &                &     & $-9.10$  & $-10.46$ \\
Pr III& $-9.91$ (0.02) & 2   & $-9.67$ (:)    & 1   & $-10.34$ (:)   & 1   & $-8.87$  & $-11.33$ \\
Nd II & $-8.72$ (0.13) & 3   & $-8.27$ (:)    & 1   & $-8.52$ (0.32) & 4   &          & $-10.59$ \\
Nd III& $-8.64$ (0.10) & 3   & $-8.54$ (0.25) & 2   & $-8.47$ (0.09) & 2   & $-8.41$  & $-10.59$ \\
Sm II & $-8.43$ (0.30) & 7   & $-8.99$ (0.74) & 2   &                &     & $-8.70$  & $-11.03$ \\
Eu II & $-8.39$ (0.27) & 3   &                &     & $-8.31$ (0.03) & 2   & $-8.85$  & $-11.52$ \\
Gd II & $-8.62$ (0.24) & 12  & $-8.39$ (0.31) & 10  & $-8.43$ (0.26) & 7   & $-8.72$  & $-10.92$ \\
\hline\hline
\end{tabular}}
\end{table}

\clearpage
\centerline {\bf ðïäðéóé ë òéóõîëáí}
\vspace{1 cm}

Fig. 1.~Spectra of the stars HD 5797 (black curve) and HD 40711 (gray curve) in the region 6144--6162\AA.

Fig. 2.~Comparison of the theoretical H$\alpha$ and H$\beta$ line profiles (black curves) with those observed in the spectra of HD 5797 (gray curves).

Fig. 3.~Positions of the stars HD 5797 and HD 40711 on the theoretical Hertzsprung-Russell diagram.

Fig. 4.~The curve of longitudinal magnetic field variability in HD 5797 constructed from our measurements in accordance with the ephemeris by Wolff  (1975).

Fig. 5.~Comparison of the equivalent widths of some Fe, Cr, and Ti lines in the spectra taken at different rotation phases.

\clearpage
\begin{figure}[h]
\epsfxsize=19cm
\hspace{-2cm}\epsffile{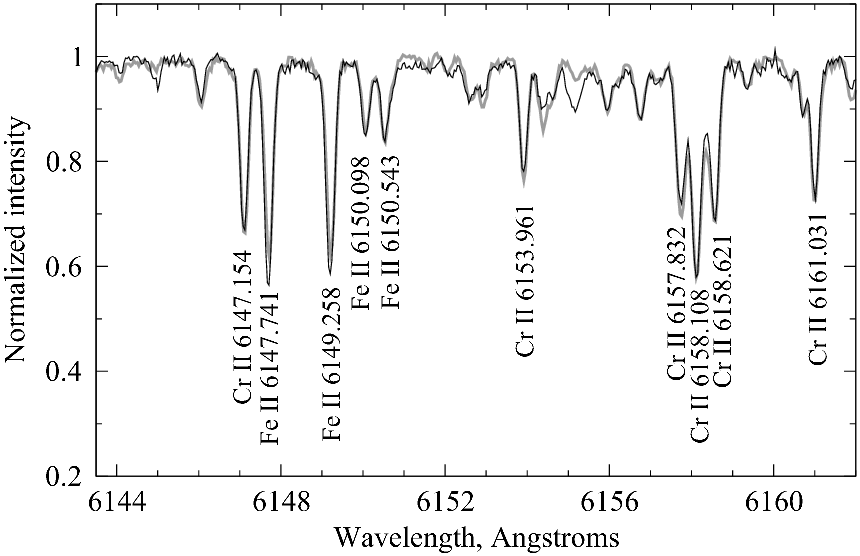}
\caption{\rm}
\label{figure:1}
\end{figure}

\clearpage
\begin{figure}[h]
\epsfxsize=19cm
\hspace{-2cm}\epsffile{fig2.eps}
\caption{\rm}
\label{figure:2}
\end{figure}

\clearpage
\begin{figure}[h]
\epsfxsize=19cm
\hspace{-2cm}\epsffile{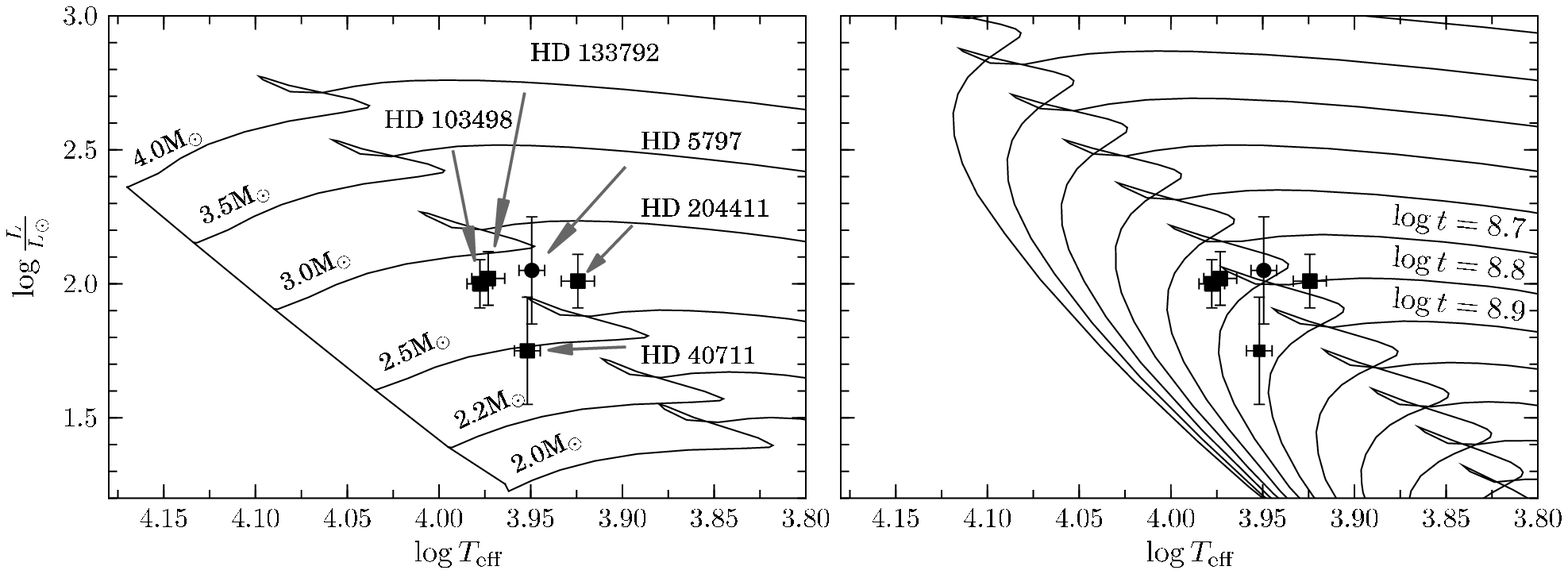}
\caption{\rm}
\label{figure:3}
\end{figure}

\clearpage
\begin{figure}[h]
\epsfxsize=19cm
\hspace{-2cm}\epsffile{fig4.eps}
\caption{\rm}
\label{figure:4}
\end{figure}

\clearpage
\begin{figure}[h]
\epsfxsize=19cm
\hspace{-2cm}\epsffile{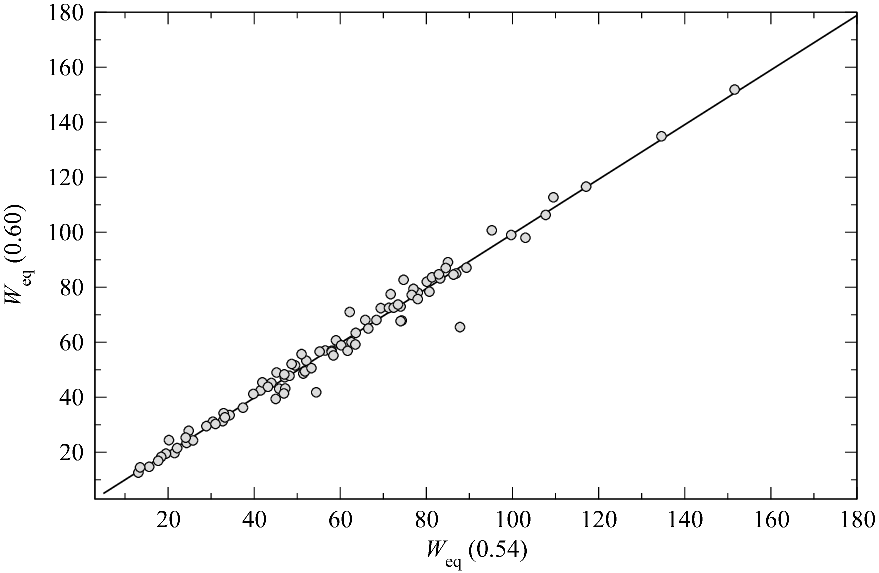}
\caption{\rm}
\label{figure:5}
\end{figure}

\end{document}